\documentclass [12pt,a4paper]{article}
\usepackage{bbm}
\usepackage{amsfonts}
\usepackage{mathrsfs}
\usepackage {amssymb}
\usepackage {amsmath}
\usepackage{amsthm}
\usepackage{latexsym}
\usepackage {amssymb}
\usepackage {amsmath}
\usepackage{color}
\usepackage{dcolumn}
\usepackage{extarrows}
\makeatletter
 \@addtoreset{equation}{section}
 
\makeatother

\usepackage{lastpage}
\usepackage{epsfig}

\begin{document}
\begin{center}
\textbf{\Large{On the 2-Adic Complexity of the Ding-Helleseth-Martinsen Binary Sequences} }\footnote {Lulu Zhang and Jun Zhang were supported by the National Natural Science Foundation of China (NSFC) under Grant 11601350, and by the Scientific Research Project of Beijing Municipal Education Commission under Grant KM201710028001. Minghui Yang was supported by NSFC under Grant 11701553. Keqin Feng was supported by NSFC under Grant 11571007 and 11471178.

Lulu Zhang and Jun Zhang are with the School of Mathematical Sciences, Capital Normal University, Beijing 100048, China, (e-mail: 840375411@qq.com; junz@cnu.edu.cn).

Minghui Yang is with the State Key Laboratory of Information Security, Institute of Information Engineering, Chinese Academy of Sciences, Beijing 100093, China, email: (yangminghui6688@163.com).

Keqin Feng is with the department of Mathematical Sciences, Tsinghua University, Beijing, 100084, China, email: (fengkq@tsinghua.edu.cn). }

\end{center}

\begin{center}
\small Lulu  Zhang, Jun Zhang, Minghui Yang and Keqin Feng
\end{center}


\noindent\textbf{Abstract}-We determine the 2-adic complexity of the Ding-Helleseth-Martinsen (DHM) binary sequences by using cyclotomic numbers of order four, ``Gauss periods" and ``quadratic Gauss sum" on finite field $\mathbb{F}_q$ and valued in $\mathbb{Z}_{2^N-1}$ where $q \equiv 5\pmod 8$ is a prime number and $N=2q$ is the period of the DHM sequences.

\noindent\textbf{keywords}-binary sequences, autocorrelation, 2-adic complexity, Ding-Helleseth-Martinsen sequences, cyclotomic number
\section{Introduction}

 \ \ \  \ \ Let
 $S=\{s_i\}_{i=0}^{\infty}$ be a binary sequence of period $N \geq 3$, $s_t\in\{0, 1\}$ $s_{N+t}=s_{t}$. The autocorrelation function of the sequence $S$ is defined by
$$A_S(\tau)=\sum_{t=0}^{N-1}(-1)^{s_{t+\tau}-s_t} \in\mathbb{Z} \ (0\leq\tau\leq N-1).$$
For $\tau=0$, $A_S(0)=N.$ Let
$$\textrm{Max} \ A_S= \textrm{max} \{|A_S(\tau)|: 1\leq\tau\leq N\}.$$
For many applications in communication, the value of Max $A_s$ is required as small as possible. It is easy to see that when $N \geq 3$, $A_S(\tau)\equiv N\pmod 4$ for all $0\leq\tau\leq N-1.$ For $N \equiv 0\pmod 4$, a binary sequence $S$ with period $N$ is called to have perfect autocorrelation if  Max $A_S=0.$ It is conjectured that the only perfect sequence is $N=4$ and $S=(0, 0, 0, 1, \ldots)$ up to (cyclic shift) equivalence. For $N \equiv 3\pmod 4$, a binary sequence $S$ with period $N$ is called to have ideal autocorrelation if $A_S(\tau)=-1$ for all $1\leq\tau\leq N-1.$ Several series of binary sequences with ideal autocorrelation have been found ($m$-sequences, Hall sequences, Paley sequences and the twin-prime sequences, see [2]). A binary sequence $S$ with period $N$ is called to have optimal autocorrelation if Max $A_S=4, 3, 2$ and 3 for $N \equiv 0, 1, 2, 3 \pmod 4$ respectively. For a list of known binary sequences with optimal autocorrelation we refer to [2].

In the application on cryptography, binary sequences, as candidates of keys in stream cipher system, are required to have big ``complexity". There are huge works on linear complexity of binary sequences. The sequences with linear complexity $n$ can be generated by a linear shift register of length $n$. Since the end of last century, the 2-adic complexity has been viewed as one of the important security criteria of sequences. The sequences with 2-adic complexity $n$ can be generated by a feedback (with carry) shift register of length $n$ ([6]).

Let $S=(s_i)_{i=0}^{\infty}$ be a binary sequence with period $N$ $(s_i\in \{0, 1\}).$ Let $S(x)=\sum_{i=0}^{N-1}s_ix^i\in \mathbb{Z}[x], d=\gcd(S(2), 2^N-1).$ The 2-adic complexity of $S$ is defined by
$$C_2(S)=\log_2(\frac{2^N-1}{d}).$$

Comparing with the linear complexity, the 2-adic complexity of binary sequences with small autocorrelation has not been fully researched. The 2-adic complexity of the binary sequences with ideal autocorrelation $(A_S(\tau)=-1$ for $1\leq\tau\leq N-1)$ has been done in [12, 14, 5]. Particularly, H. Hu [5] presented a neat approach to show that for all known ideal binary sequences with period $N$, their 2-adic complexity reaches the maximum value $\log_2(2^N-1).$ For some other sequences with good autocorrelation, the 2-adic complexity is determined or estimated by a nice lower bound [4, 9-11, 13].

In this paper, we determine the 2-adic complexity of the Ding-Helleseth-Martinsen (DHM) binary sequences. The sequences has period $N=2q$ where $q \equiv 5 \pmod 8$ is a prime number, and optimal autocorrelation Max $A_S=2.$ We will determine their 2-adic complexity. For doing this we use the cyclotomic numbers of order four and develop ``Gauss periods" and quadratic ``Gauss sum" on finite field $\mathbb{F}_q$ valued in the ring $\mathbb{Z}_{2^N-1}$.

We introduce the construction on the DHM sequences and preliminaries on cyclotomic numbers, ``Gauss periods" and quadratic ``Gauss sum" in Section 2.
Then we present upper and lower bounds on the 2-adic complexity of the DHM sequences in Section 3. After further consideration we finally determine the exact value of the 2-adic complexity of the DHM sequences in Section 4.

\section{Preliminaries}

\subsection{Ding-Helleseth-Martinsen (DHM) sequences}

Let $q$ be a prime number, $q\equiv 5\pmod  8$. It is well known that there exists unique $s, t\in \mathbb{Z}$ up to their signs such that
$q=s^2+4t^2$. Let $\mathbb{F}_q^{\ast}=\langle\theta\rangle, q-1=4f$ and $C=\langle\theta^4\rangle$. Then $C$ is a subgroup of $\mathbb{F}_q^{\ast}$, and the cosets of $C$ in $\mathbb{F}_q^{\ast}$
\begin{equation}
D_\lambda=\theta^\lambda C=\{\theta^{\lambda+4i}: 0\leq i\leq f-1\}
\end{equation}
are called as cyclotomic classes of order four in $\mathbb{F}_q$.

Let $N=2q(\equiv 2\pmod 4).$ We have the following isomorphism of rings
$$\varphi: \mathbb{Z}_2\times \mathbb{Z}_q\cong \mathbb{Z}_{2q}=\mathbb{Z}_{n},  (a,b)\rightarrow (q+1)b+qa.$$
The inverse of $\varphi$ is $\varphi^{-1}(i)=(i\pmod 2, i\pmod q).$

Consider the following subset of $\mathbb{Z}_2\times\mathbb{Z}_q$, for $i, j, l\in\{0, 1, 2, 3\}$
\begin{equation}
C'(i, j, l)=[\{0\}\times(D_i\cup D_j)]\cup[\{1\}\times(D_l\cup D_j)], C(i, j, l)=\varphi (C'(i, j, l))
\end{equation}

$\widetilde{C}'(i, j, l)=C'(i, j, l)\cup\{(0,0)\}, \widetilde{C}(i, j, l)=\varphi(\widetilde{C}'(i, j, l))=C(i, j, l)\cup\{0\}$ \ \ \ \ \ \ \  $(\widetilde{2.2})$

\par\noindent\textbf{Definition 2.1} Let $q$ be a prime number, $q \equiv 5\pmod 8, q-1=4f, D_\lambda(\lambda=0, 1, 2, 3)$ be the cyclotomic classes of order four defined by (2.1). Let $C(i, j, l)$ be the subset of $\mathbb{Z}_{2q}$ defined by (2.2). The Ding-Helleseth-Martinsen (DHM) binary sequence $S=S(i, j, l)=\{s_\lambda\}_{\lambda=0}^{\infty}$ and $\widetilde{S}=\widetilde{S}(i, j, l)=\{\widetilde{s}_\lambda\}_{\lambda=0}^{\infty}$ with period $N=2q$ is defined by
\begin{align*}
 s_\lambda
 = \left\{ \begin{array}{ll}
1, & \textrm{if $\lambda\in C(i, j, l) (\textrm{Namely}, \lambda=(q+1)b+qa$}) \\
   & \textrm{{where} ``$a=0, b\in D_i\cup D_j"$ \textrm{or} ``$a=1, b\in D_l\cup D_j"$ }\\
0, & \textrm{otherwise},
\end{array} \right.
\end{align*}

\begin{align*}
\widetilde{s}_\lambda
 = \left\{ \begin{array}{ll}
1, & \textrm{if $\lambda\in\widetilde{C}(i, j, l),$}\\
0, & \textrm{otherwise}.
\end{array} \right.
\end{align*}
Namely, $\widetilde{s}_0=1$ and $\widetilde{s}_\lambda=s_\lambda$ for $1\leq\lambda\leq N-1$.

It is proved in [3] that if
\begin{align}
   \begin{array}{ll}
& \textrm{(I)}\ t=1\ \textrm{and}\ (i, j, l)=(0, 1, 3), (0, 2, 1); \textrm{or}\\
& \textrm{(II)}\ s=1 \ \textrm{and}\ (i, j, l)=(1, 0, 3), (0, 1, 2),
\end{array}
\end{align}
then the DHM sequence $S=S(i, j, l)$ has optimal autocorrelation $(A_S(\tau)=\pm 2$ for $1\leq \tau\leq N-1=2q-1).$

If
\begin{align*}
   \begin{array}{ll}
& \textrm{$\rm(\widetilde{I})$}\ t=1\ \textrm{and}\ (i, j, l)=(0, 1, 3), (0, 2, 3), (1, 2, 0), (1, 3, 0) \ \textrm{or}\\
& \textrm{$\rm(\widetilde{II})$}\ s=1 \ \textrm{and}\ (i, j, l)=(0, 1, 2), (0, 3, 2), (1, 0, 3), (1, 2, 3).\ \ \ \ \ \ \ \ \ \ \ \  \ \ \ \ \ \ \ \ \ \ \  \ \ \ \ (\widetilde{2.3})
\end{array}
\end{align*}
then the DHM sequences $\widetilde{S}=\widetilde{S}(i, j, l)$ has optimal autocorrelations.

In Section 3 we present upper and lower bounds for the DHM sequence $S=S(i, j, l)$ satisfying condition (2.3) and $\widetilde{S}=\widetilde{S}(i, j, l)$ satisfying condition $(\widetilde{2.3})$ (Theorem 3.1 and 3.2). After further consideration we totally determine the value of $C_2(S)$ and $\widetilde{C}_2(S)$ in Section 4 (Theorem 4.2 and 4.4).\\

\subsection{Cyclotomic Numbers of Order Four}

Let $q\equiv 5\pmod 8$ be a prime number, $\mathbb{F}_q^{\ast}=\langle\theta\rangle$, $C=\langle\theta^4\rangle$ and $D_\lambda=\theta^\lambda C \ (0\leq\lambda\leq3)$ be the cyclotomic classes of order four in $\mathbb{F}_q$.

  \par\noindent\textbf{Definition 2.2} The cyclotomic numbers of order four in $\mathbb{F}_q$ are defined by, for $0\leq i,j\leq 3$
  $$(i, j)=|(D_i+1)\cap D_j|=\sharp \{(a,b): a\in D_i, b\in D_j, a+1=b\}.$$
  The values of $(i,j)$ has been computed (see [1] or [8]).

  \par\noindent\textbf{Lemma 2.3} Let $q\equiv 5\pmod 8$ be a prime number, $q=s^2+4t^2, s, t\in\mathbb{Z}$ and $s\equiv 1\pmod 4$. The values of the cyclotomic numbers $(i, j)$ of order four in $\mathbb{F}_q$ are
  $$16(0,0)=16(2,2)=16(2,0)=A=q-7+2s,$$
  $$16(0,1)=16(1,3)=16(3,2)=B=q+1+2s-8t,$$
  $$16(1,2)=16(0,3)=16(3,1)=\bar{B}=q+1+2s+8t,$$
  $$16(0,2)=C=q+1-6s,$$
  $$16(1,0)=16(1,1)=16(2,1)=16(3,0)=16(3,3)=16(2,3)=D=q-3-2s.$$

\subsection{``Gauss periods" of order four and quadratic ``Gauss sum"}

 Since $a\equiv b\pmod q$ implies $4^a\equiv 4^b \bmod (2^N-1) (N=2q),$ we can define the following mapping
$$f: \mathbb{Z}_q\rightarrow \mathbb{Z}_{2^N-1}^{\ast}, f(a)= 4^a$$
where $\mathbb{Z}_m^{\ast}$ is the group of units in the ring $\mathbb{Z}_m=\mathbb{Z}/m\mathbb{Z} \ (m\geq 2)$. $f$ is a homomorphism of groups from $(\mathbb{Z}_q, +)$ to $(\mathbb{Z}_{2^N-1}^{\ast}, \cdot)$, and can be viewed as an additive character of finite field $\mathbb{Z}_q=\mathbb{F}_q$ valued in $\mathbb{Z}_{2^N-1}^{\ast}$. Then we have the ``Gauss periods" of order 4
$$\eta_\lambda=\sum_{i\in D_\lambda}4^i\bmod (2^N-1) (\lambda=0, 1, 2, 3)$$
and quadratic ``Gauss sum"
$$G=\sum_{i\in \mathbb{F}_q^{\ast}}4^i\chi(i)=\eta_0-\eta_1+\eta_2-\eta_3\ (\bmod \ 2^N-1),$$
where $\chi$ is the unique quadratic (multiplicative) character of $\mathbb{F}_q^{\ast}$ (the Legendre symbol). Namely, for $i\in \mathbb{F}_q^{\ast}$,

\begin{align*}
 \chi(i)
 = \left\{ \begin{array}{ll}
1, & \textrm{if $i\in D_0\cup D_2=\langle\theta^2\rangle$}\\
-1, & \textrm{if $i\in D_1\cup D_3=\theta\langle\theta^2\rangle$}.
\end{array} \right.
\end{align*}

The following results show that $\eta_\lambda$ and $G$ have some similar properties as usual Gauss periods and Gauss sums. (The proofs are also similar).
\par\noindent\textbf{Lemma 2.4} (1). $G^2\equiv q-\frac{4^q-1}{3}\ (\bmod \ 4^q-1)$.\\
 (2). For $0\leq\lambda,\mu\leq3,$
 $\eta_\lambda\eta_\mu\equiv\frac{q-1}{4}\delta_{\lambda, \mu+2}+\sum_{\nu=0}^3(\lambda-\nu+2, \mu-\nu)\eta_\nu \ (\bmod \ 4^q-1)$,
 where $(a, b)$ is the cyclotomic numbers of order four on $\mathbb{F}_q$ and
\begin{align*}
 \delta_{\lambda, \mu}
 = \left\{ \begin{array}{ll}
1, & \textrm{if $\lambda\equiv\mu \pmod 4$}\\
0, & \textrm{otherwise}.
\end{array} \right.
\end{align*}
\begin{proof}
(1). In the following, all equality $a=b$ means $a\equiv b \ (\bmod \ 4^q-1).$
\begin{align}
G^2& =\sum_{a,b=1}^{q-1}4^{a+b}\chi(ab)=\sum_{a,c=1}^{q-1}4^{a(1+c)}\chi(a^2c) \ (\textrm{taking}\ c=ba^{-1})\notag\\
                                  &= \sum_{c=1}^{q-1}\chi(c)\sum_{a=1}^{q-1}4^{a(1+c)}\notag.
                                 \end{align}
The contribution of $c=q-1$ to the right-hand side is
$$\chi(q-1)\sum_{a=1}^{q-1}4^{aq}=\chi(-1)\sum_{a=1}^{q-1}1=q-1$$
(remark that $\chi(-1)=1$ for $q\equiv 5\equiv1\pmod 4).$ Therefore
$$G^2=q-1+\sum_{c=1}^{q-2}\chi(c)(-1+\sum_{a=0}^{q-1}4^a)=q-\sum_{a=0}^{q-1}4^a=q-\frac{4^q-1}{3}.$$
(2). By the definition of $\eta_\lambda$,
  \begin{align}
\eta_\lambda\eta_\mu& =\sum_{a\in D_\lambda\atop b\in D_\mu}4^{a+b}=\sum_{c\in\mathbb{F}_q}4^c\sum_{a\in D_\lambda\atop c-a\in D_\mu}1 \ (c=a+b)\notag\\
                                  &= \sum_{a\in D_\lambda\atop -a\in D_\mu}1+\sum_{\nu=0}^3\sum_{c\in D_\nu}4^c\sum_{e\in D_{\lambda-v+2}\atop 1+e\in D_{\mu-\nu}}1 \ ( \textrm{let}\ e=-\frac{a}{c}\ \textrm{for}\ c\neq 0)\notag.
                                 \end{align}
From $q \equiv 5\pmod 8$  we know that $\frac{q-1}{2}\equiv 2 \pmod 4$ and $-1=\theta^{\frac{q-1}{2}}\in D_2.$\\
Therefore
$$a\in D_\lambda\Longleftrightarrow e(=-\frac{a}{c})\in D_{\lambda-\nu+2}$$
$$c-a\in D_\mu\Longleftrightarrow 1+e(=\frac{1}{c}(c-a))\in D_{\mu-\nu}$$
Then we get
$$\eta_\lambda\eta_\mu=\frac{q-1}{4}\delta_{\lambda, \mu+2}+\sum_{\nu=0}^3(\lambda-\nu+2, \mu-\nu)\eta_\nu.$$
\end{proof}
 \par\noindent\textbf{Remark} From Lemma 2.4(2) we know that
 \begin{align}
\eta_{\lambda+i}\eta_{\mu+i}& =\frac{q-1}{4}\delta_{\lambda+i, \mu+i+2}+\sum_{\nu=0}^3(\lambda+i-\nu+2, \mu+i-\nu)\eta_\nu\notag\\
                                  &= \frac{q-1}{4}\delta_{\lambda, \mu+2}+\sum_{\nu=0}^3(\lambda-\nu+2, \mu-\nu)\eta_{\nu+i}.
                                 \end{align}
By Lemma 2.3 we get
\begin{align}
   \begin{array}{ll}
& 16\eta_0^2=A\eta_0+B\eta_1+C\eta_2+\overline{B}\eta_3,\ 16\eta_0\eta_1=D\eta_0+D\eta_1+\overline{B}\eta_2+B\eta_3\\
& 16\eta_0\eta_2=\frac{q-1}{4}+A\eta_0+D\eta_1+A\eta_2+D\eta_3,\ 16\eta_0\eta_3=D\eta_0+\overline{B}\eta_1+B\eta_2+D\eta_3
\end{array}
\end{align}
From (2.4) and (2.5) we get
 \par\noindent\textbf{Lemma 2.5} Let $q\equiv 5\pmod 8$ be a prime number, $q=s^2+4t^2,$ $s\equiv1\pmod 4.$ Then
 \begin{align*}
16\eta_\lambda^2& =A\eta_\lambda+B\eta_{\lambda+1}+C\eta_{\lambda+2}+\overline{B}\eta_{\lambda+3}\\
                                  &= q(\eta_0+\eta_1+\eta_2+\eta_3)+(-7+2s)\eta_\lambda+(1+2s-8t)\eta_{\lambda+1}\\ & \ \ \ +(1-6s)\eta_{\lambda+2}+(1+2s+8t)\eta_{\lambda+3}\notag\\
                                  &=(\frac{4^q-1}{3}-1)q+(-7+2s)\eta_\lambda+(1+2s-8t)\eta_{\lambda+1}+(1-6s)\eta_{\lambda+2}+(1+2s+8t)\eta_{\lambda+3}
                                 \end{align*}
\begin{equation*}
16\eta_{\lambda}\eta_{\lambda+1}=(\frac{4^q-1}{3}-1)q+(-3-2s)(\eta_\lambda+\eta_{\lambda+1})+(1+2s+8t)\eta_{\lambda+2}+(1+2s-8t)\eta_{\lambda+3}
\end{equation*}
\begin{equation*}
16\eta_{\lambda}\eta_{\lambda+2}=(\frac{4^q-1}{3}-1)q+(-7+2s)(\eta_\lambda+\eta_{\lambda+2})+(-3-2s)(\eta_{\lambda+1}+\eta_{\lambda+3})+4(q-1)
\end{equation*}
\begin{equation*}
16\eta_{\lambda}\eta_{\lambda+3}=16\eta_{\lambda+3}\eta_{(\lambda+3)+1}.
\end{equation*}
 \par\noindent\textbf{Lemma 2.6} Assume that $q\equiv 5\pmod 8$ is a prime number and $q=s^2+4t^2,$ $s\equiv 1\pmod 4$. Then
 $$(\eta_0-\eta_1)^2+ (\eta_2-\eta_3)^2=-tG,\ 2(\eta_0-\eta_2)^2=-(sG+q)+\frac{4^q-1}{3}$$
  $$(\eta_0-\eta_3)^2+ (\eta_2-\eta_1)^2=tG,\ 2(\eta_1-\eta_3)^2=sG-q+\frac{4^q-1}{3}$$
  \begin{proof}
  By Lemma 2.5,
   \begin{align*}
16[(\eta_0-\eta_1)^2+(\eta_2-\eta_3)^2]& =16(\eta_0^2+\eta_1^2+\eta_2^2+\eta_3^2-2\eta_0\eta_1-2\eta_2\eta_3)\\
                                  &= 0(\eta_0+\eta_1+\eta_2+\eta_3)-16t(\eta_0-\eta_1+\eta_2-\eta_3)\notag\\
                                  &=-16tG.
                                 \end{align*}
We get $(\eta_0-\eta_1)^2+ (\eta_2-\eta_3)^2=-tG$.

Similarly, we get
$$(\eta_0-\eta_3)^2+(\eta_2-\eta_1)^2=tG  $$

Since $\eta_0+\eta_1+\eta_2+\eta_3=\frac{4^q-1}{3}-1$, by Lemma 2.5, we get  $$2(\eta_0-\eta_2)^2=-(sG+q)+\frac{4^q-1}{3},
  \ 2(\eta_1-\eta_3)^2=sG-q+\frac{4^q-1}{3}.$$
\end{proof}

\section{ Upper and Lower Bounds of $C_2(S)$ and $C_2(\widetilde{S})$}
 In this section we present upper and lower bounds of 2-adic complexity of DHM sequences.
\par\noindent\textbf{Theorem 3.1} Let $S=S(i, j, l)=\{s_\lambda\}_{\lambda=0}^{\infty}$ be the DHM binary sequence with period $N=2q$ defined in Definition 2.1, $q\equiv 5\pmod 8$ be a prime number, $q=s^2+4t^2, s\equiv 1\pmod 4.$ If the condition (2.3) holds, then $\log_2(\frac{2^N-1}{3})\geq C_2(S)\geq \log_2(\frac{2^N-1}{3D})$, where $D=\gcd(2^N-1, q^2+3q+4).$
\begin{proof}
The 2-adic complexity of $S$ is $C_2(S)=\log_2(\frac{2^N-1}{d})$, where $d=\gcd(S(2), 2^N-1)$, and $S(2)=\sum_{\lambda=0}^{N-1}s_{\lambda}2^\lambda$.
We need to estimate the value $d$. From $2^N-1=(2^q-1)(2^q+1)$ and $\gcd(2^q-1, 2^q+1)=1$, we know that
$$d=d_1d_2, d_1=\gcd(S(2), 2^q-1), d_2=\gcd(S(2), 2^q+1).$$
By the definition of the sequence $S$, we have
\begin{align*}
S(2)& =\sum_{\lambda=0}^{N-1}s_\lambda2^\lambda\equiv\sum_{b\in D_i\cup D_j}2^{(q+1)b}+\sum_{b\in D_l\cup D_j}2^{q+(q+1)b}\ (\bmod \ 2^N-1)\
                                 \end{align*}
Let $2\in D_k$. From $q\equiv 5\pmod 8$ we know that 2 is not a square in $\mathbb{F}_q^{\ast}.$ Therefore $k=1$ or 3 and
\begin{align}
S(2)& \equiv\sum_{2b\in D_i\cup D_j}2^{(q+1)2b}+\sum_{2b\in D_l\cup D_j}2^{q+(q+1)2b}\ (\bmod \ 2^N-1)\notag\\
                                  &\equiv \sum_{b\in D_{i-k}\cup D_{j-k}}4^b+2^q\sum_{b\in D_{l-k}\cup D_{j-k}}4^b\ (\bmod \ 2^N-1)\ \ (N=2q) \notag\\
                                   &\equiv \eta_{i-k}+\eta_{j-k}+2^q(\eta_{l-k}+\eta_{j-k}) \ (\bmod \ 2^N-1)
                                 \end{align}

(A). Firstly we prove $d_2=3$.

The assumption $q\equiv 5\equiv 1\pmod 2$ implies $3|2^q+1.$ On the other hand,
$$\eta_\lambda=\sum_{a\in D_\lambda}4^a\equiv\sum_{a\in D_\lambda}1=\frac{q-1}{4} \pmod 3 (\lambda\in \{0, 1, 2, 3\}),$$
we get $S(2)=\eta_{i-k}+\eta_{j-k}-\eta_{l-k}-\eta_{j-k}\equiv 0\pmod 3.$ Therefore $3|d_2=\gcd(S(2), 2^q+1).$ Moreover, if $9|2^q+1$, then $2^q\equiv -1\pmod 9$. The order of $2 \pmod 9$ is 6, we get $q=3+6l$ which implies that $q=3$ since $q$ is a prime number. This contradicts to assumption $q\equiv 5\pmod 8.$ Thus $9\nmid 2^q+1$ and $d_2=3d'_2$ where $3\nmid d'_2=\gcd(S(2), \frac{2^q+1}{3}).$

Now we prove $d'_2=1.$ By (3.1),
\begin{align}
0& \equiv S(2)\equiv \eta_{i-k}+\eta_{j-k}-\eta_{l-k}-\eta_{j-k}\equiv \eta_{i-k}-\eta_{l-k}\ (\bmod \ 2^q+1)\notag\\
                                  &\equiv \left\{ \begin{array}{ll}
\eta_{0-k}-\eta_{3-k} & \\
\eta_{0-k}-\eta_{1-k} & \\
\eta_{1-k}-\eta_{3-k} & \\
\eta_{0-k}-\eta_{2-k}
\end{array} \right.
\pmod {d'_2} \ \textrm{if} \ (i, j, l)\ = \left\{ \begin{array}{ll}
(0, 1, 3)& \\
(0, 2, 1) & \\
(1, 0, 3) & \\
(0, 1, 2)
\end{array} \right.
\end{align}
where $k=1$ or 3.
We consider the four cases of the condition (2.3) separately.

(1). For $t=1$ and $(i, j, l)=(0, 1, 3)$, we have $\eta_{-k}-\eta_{3-k}\equiv 0\pmod {d'_2}$ by (3.2) and $k=1$ or 3. Namely, $\eta_3-\eta_2\equiv 0$ or $\eta_1-\eta_0\equiv 0\pmod {d'_2}$. Then
$G=\eta_0-\eta_1+\eta_2-\eta_3\equiv -(\eta_1-\eta_0)$ or $\eta_2-\eta_3\pmod {d'_2}$. In both cases,
\begin{align*}
G^2& \equiv (\eta_1-\eta_0)^2+(\eta_2-\eta_3)^2\pmod {d'_2} \\
                                  &\equiv -G\pmod {d'_2} \ \textrm{(by Lemma 2.6 and $t=1$)}
                                 \end{align*}
From Lemma 2.4 we have
\begin{align}
G^2& \equiv q-\frac{4^q-1}{3}\pmod {d'_2} \notag\\
                                  &\equiv q\pmod {d'_2}\ \textrm{(since $3\nmid {d'_2}$ and ${d'_2}| 4^q-1$)}.
                                 \end{align}
Therefore $-G\equiv q$ and $q\equiv G^2 \equiv q^2 \pmod {d'_2}$. We get $d'_2|\gcd(q^2-q, \frac{2^q+1}{3}).$

(2). For $t=1$ and $(i, j, l)=(0, 2, 1)$, we have $\eta_{-k}-\eta_{1-k}\equiv 0\pmod {d'_2}$. Namely, $0\equiv\eta_3-\eta_0$ or $\eta_1-\eta_2\pmod {d'_2}$ and $G\equiv -\eta_1+\eta_2$ or $\eta_0-\eta_3\pmod {d'_2}$. In both of cases,
\begin{align*}
q&\equiv G^2\equiv(\eta_0-\eta_3)^2+(\eta_1-\eta_2)^2
                                  \equiv G\pmod {d'_2}\ \textrm{(by Lemma 2.6 and $t=1$)}.
                                 \end{align*}
Therefore $q \equiv G^2\equiv q^2\pmod {d'_2}$ and we also get $d'_2|\gcd(q^2-q, \frac{2^q+1}{3})$ as in case (1).

(3). For $s=1$ and $(i, j, l)=(1, 0, 3),$ we have $\eta_{1-k}-\eta_{3-k}\equiv0\pmod {d'_2}$. Namely, $0=\eta_0-\eta_2$ or $\eta_2-\eta_0 \pmod {d'_2}.$ Therefore
\begin{align*}
0&\equiv 2(\eta_0-\eta_2)^2\equiv-(G+q)+\frac{4^q-1}{3}\pmod {d'_2} \ \textrm{(by Lemma 2.6 and $s=1$)}
 \end{align*}
From $3\nmid {d'_2}$ we get $\frac{4^q-1}{3}\equiv 0\pmod {d'_2}$.
Therefore $G\equiv -q$ and $q\equiv G^2\equiv q^2\pmod {d'_2}.$ Then we get $d'_2|\gcd(q^2-q, \frac{2^q+1}{3})$.

(4). At last, for $s=1$ and $(i, j, l)=(0, 1, 2)$, we have $\eta_{-k}-\eta_{2-k}\equiv 0\pmod {d'_2}$. Namely, $0\equiv \eta_3-\eta_1$ or $\eta_1-\eta_3 \pmod {d'_2}$.
Therefore
\begin{align*}
0&\equiv 2(\eta_1-\eta_3)^2
                                  \equiv G-q+\frac{4^q-1}{3}\pmod {d'_2}\ \textrm{(by Lemma 2.6 and $s=1$)}.
                                 \end{align*}
We also get $d'_2|\gcd(q^2-q, \frac{2^q+1}{3})$.

In summary, we have proved that $d'_2|\gcd(q^2-q, \frac{2^q+1}{3})$ for all four cases.

Now we prove that $g=\gcd(q^2-q, \frac{2^q+1}{3})=1$. If $g>1$, let $p$ be a prime divisor of $g$. Then $p\geq 5$, $p=q$ or $p|q-1$. If $p=q$, we get $2^q\equiv 2\pmod p$ which contradicts to $p|2^q+1$. If $p|q-1$ from $2^q\equiv -1\pmod p$ we know that the order of $4\pmod p$ is $q$. Therefore $q|p-1$ which contradicts to $p|q-1$. Therefore $g=1, d'_2=1$ and then $d_2=3$ for all four cases.

This implies that $C_2(S)\leq \log_2(\frac{2^N-1}{3})$.

(B). Now we prove that $d_1|\gcd(q^2+3q+4, 2^q-1)$.

From (3.1) we know that
\begin{align*}
 S(2)& \equiv \eta_{i-k}+\eta_{j-k}+\eta_{l-k}+\eta_{j-k} \ (\bmod \ 2^q-1)\notag\\
                                  &\equiv -1+\eta_{j-k}-\eta_{\lambda-k} \ (
                                   \bmod \ d_1)\ \textrm{(where} \ \lambda\in\{0, 1, 2, 3\}\backslash \{i, j, l\})\notag
                                  \end{align*}
Therefore
\begin{align}
 0 \equiv S(2)&\equiv \left\{ \begin{array}{ll}
-1+\eta_{1-k}-\eta_{2-k} & \\
-1+\eta_{2-k}-\eta_{3-k} & \\
-1+\eta_{-k}-\eta_{2-k} & \\
-1+\eta_{1-k}-\eta_{3-k}
\end{array} \right.
\ (\bmod \ d_1) \ \textrm{if} \ (i, j, l)\ = \left\{ \begin{array}{ll}
(0, 1, 3)& \\
(0, 2, 1) & \\
(1, 0, 3) & \\
(0, 1, 2)
\end{array} \right.
\end{align}

Since $d_1|(2^q-1)$, we have $\frac{4^q-1}{3}\equiv 0\ (\bmod \ d_1)$ and then $\eta_0+\eta_1+\eta_2+\eta_3=\frac{4^q-1}{3}-1\equiv -1\ (\bmod \ d_1)$.

(1). For $t=1$ and $(i, j, l)=(0, 1, 3)$, we have $1\equiv\eta_0-\eta_1$  or $\eta_2-\eta_3(\bmod \ d_1)$ and $1-G \equiv -\eta_2+\eta_3$ or $-\eta_0+\eta_1\ (\bmod \ d_1)$.
Therefore by Lemma 2.6 and $t=1,$
$$1^2+(1-G)^2\equiv (\eta_0-\eta_1)^2+(\eta_2-\eta_3)^2\equiv -G \ (\bmod \ d_1)$$
Since $1^2+(1-G)^2=G^2-2G+2\equiv q-2G+2 \ (\bmod \ d_1)$, we get $G\equiv q+2$ and $q\equiv G^2\equiv q^2+4q+4 \ (\bmod \ d_1)$. Therefore $q^2+3q+4\equiv 0\ (\bmod \ d_1)$ and then
$d_1|\gcd(q^2+3q+4, 2^q-1).$

(2). For $t=1$ and $(i, j, l)=(0, 2, 1)$, we get $1\equiv \eta_1-\eta_2$ or $\eta_3-\eta_0 \ (\bmod \ d_1)$ and $1+G\equiv \eta_0-\eta_3$ or $\eta_2-\eta_1\ (\bmod \ d_1)$. Then by Lemma 2.6 we have
$$1+(1+G)^2\equiv(\eta_0-\eta_3)^2+(\eta_1-\eta_2)^2\equiv G\ (\bmod \ d_1).$$ Since $1+(1+G)^2=q+2G+2,$ we get $G\equiv -q-2$ and $q\equiv G^2\equiv q^2+4q+4\ (\bmod \ d_1)$
Then we also get $d_1|\gcd(q^2+3q+4, 2^q-1).$

(3).  For $s=1$ and $(i, j, l)=(1, 0, 3)$, we get $1\equiv \eta_3-\eta_1$ or $\eta_1-\eta_3\ (\bmod \ d_1)$. By Lemma 2.6 and $s=1$, $2\equiv 2(\eta_1-\eta_3)^2\equiv G-q$ and then $G\equiv q+2 \ (\bmod \ d_1) $. We also get
 $d_1|\gcd(q^2+3q+4, 2^q-1).$

(4). At last, for $s=1$ and $(i, j, l)=(0, 1, 2)$, we get $1\equiv\eta_0-\eta_2$ or $\eta_2-\eta_0\ (\bmod\ d_1)$. By Lemma 2.6 and $s=1$ we have  $2\equiv 2(\eta_0-\eta_2)^2\equiv -G-q$ and $G\equiv -q-2\ (\bmod \ d_1)$. We also get $d_1|\gcd(q^2+3q+4, 2^q-1).$

In summary, we have proved $d_1|\gcd(q^2+3q+4, 2^q-1)$ for all four cases. This completes the proof of $C_2(S)\geq \log_2(\frac{2^N-1}{3D})$ for $N=2q$ and $D=\gcd(q^2+3q+4, 2^q-1).$
\end{proof}

\par\noindent\textbf{Theorem 3.2} Let $\widetilde{S}=\widetilde{S}(i, j, l)$ be the DHM binary sequence with $N=2q$ defined in Definition 2.1, $q\equiv 5\pmod 8$ be a prime number, $q=s^2+4t^2$, $s\equiv 1\pmod 4$. If the condition $(\widetilde{2.3})$ holds, then
$$C_2(\widetilde{S})\geq \log_2(\frac{2^N-1}{D}), D=\gcd(q^2+3q+4, 2^q+1).$$
\begin{proof}
The 2-adic complexity of $\widetilde{S}$ is $C_2(\widetilde{S})=\log_2(\frac{2^N-1}{d})$ where $d=\gcd(\widetilde{S}(2), 2^N-1)$ and $\widetilde{S}(2)=S(2)+1$. We need to estimate the value of $d$. Let $d_1=\gcd(\widetilde{S}(2), 2^q-1)$, $d_2=\gcd(\widetilde{S}(2), 2^q+1)$. Then
$d=d_1d_2$. We have proved $S(2)\equiv 0\pmod 3$ in the proof of Theorem 3.1. Thus $3\nmid \widetilde{S}(2)=S(2)+1$ and then $3\nmid d$.

(A). First we show that $d_1=1$.

(1) Suppose that $s=1$. From $\widetilde{S}(2)=1+S(2)\equiv \eta_{j-k}-\eta_{\lambda-k}\pmod {2^q-1} (k=1 \ \textrm{or} \ 3, \lambda\in\{0, 1, 2, 3\}\backslash \{i, j, l\}$) we get \begin{align*}
 0 \equiv \widetilde{S}(2)&\equiv \left\{ \begin{array}{ll}
\eta_{1-k}-\eta_{3-k} & \\
\eta_{3-k}-\eta_{1-k} & \\
\eta_{-k}-\eta_{2-k} & \\
\eta_{2-k}-\eta_{-k}
\end{array} \right.
\ (\bmod \ d_1) \ \textrm{if} \ (i, j, l)\ = \left\{ \begin{array}{ll}
(0, 1, 2)& \\
(0, 3, 2) & \\
(1, 0, 3) & \\
(1, 2, 3)
\end{array} \right.
\end{align*}
For $(i, j, l)=(0, 1, 2)$, $0\equiv \eta_0-\eta_2$ or $\eta_2-\eta_0\pmod {d_1}$ and then
$$0\equiv 2(\eta_0-\eta_2)^2\equiv -G-q \pmod {d_1}\ \ \textrm{(by Lemma 2.6)}$$
Thus $q\equiv G^2\equiv q^2\pmod {d_1}$ which implies that $d_1|\gcd(q^2-q, 2^q-1)$.\\
For $(i, j, l)=(0, 3, 2), 0\equiv \eta_2-\eta_0$ or $\eta_0-\eta_2\pmod {d_1}$ we also get $d_1|\gcd(q^2-q, 2^q-1)$.\\
For $(i, j, l)=(1, 0, 3), 0\equiv \eta_1-\eta_3$ or $\eta_3-\eta_1\pmod {d_1}$ and then
$$0\equiv 2(\eta_1-\eta_3)^2\equiv G-q\pmod {d_1}.$$
We also get $G\equiv q$ and $q\equiv 1\pmod {d_1}$. Therefore $d_1|\gcd(q^2-q, 2^q-1)$.\\
For $(i, j, l)=(1, 2, 3), 0\equiv \eta_3-\eta_1$ or $\eta_1-\eta_3\pmod {d_1}$. We also have $d_1|\gcd(q^2-q, 2^q-1)$.

If $d_1>1$, let $l$ be a prime divisor of $d_1$. Then $l|q^2-q$ and $l|2^q-1$. If $l=q$, then $0\equiv 2^q-1\equiv 2-1\equiv 1\pmod q$,
a contradiction. If $l|q-1$, then $2^q\equiv 1\pmod l$ implies that the order of $2\pmod l$ is $q$. Then $q|l-1$ which contradicts to $l|q-1$. Therefore we get $d_1=1$.

(2). Assume that $t=1$. From $\widetilde{S}(2)=S(2)+1\equiv\eta_{j-k}-\eta_{\lambda-k}\pmod{2^q-1}$ we get
\begin{align*}
 0 \equiv \widetilde{S}(2)&\equiv \left\{ \begin{array}{ll}
\eta_{1-k}-\eta_{2-k} & \\
\eta_{2-k}-\eta_{1-k} & \\
\eta_{2-k}-\eta_{3-k} & \\
\eta_{3-k}-\eta_{2-k}
\end{array} \right.
\ (\bmod \ d_1) \ \textrm{if} \ (i, j, l)\ = \left\{ \begin{array}{ll}
(0, 1, 3)& \\
(0, 2, 3) & \\
(1, 2, 0) & \\
(1, 3, 0)
\end{array} \right.
\end{align*}
For $(i, j, l)=(0, 1, 3)$, we get

$0\equiv \eta_0-\eta_1$ (or  $\eta_2-\eta_3$), $G\equiv \eta_2-\eta_3$ (or $\eta_0-\eta_1)\pmod{d_1}.$
Therefore by Lemma 2.6,
$$q\equiv G^2\equiv (\eta_0-\eta_1)^2+(\eta_2-\eta_3)^2\equiv -G\pmod{d_1}$$
which implies $q^2\equiv G^2\equiv q\pmod{d_1}$ and then $d_1=1$ as before. Similarly, for $(i, j, l)$=(0, 2, 3), (1, 2, 0) and (1, 3, 0) we get
$G\equiv\pm q\pmod{d_1}$ and then $q\equiv G^2\equiv q^2\pmod{d_1}$. We also get $d_1=1.$

(B). Now we prove that $d_2|\gcd(q^2+3q+4, 2^q+1).$

In this case $\widetilde{S}(2)=S(2)+1\equiv 1+\eta_{i-k}-\eta_{l-k}\pmod{2^q+1}$.

(1) Assume that $s=1$, we get
\begin{align*}
 0 \equiv \widetilde{S}(2)&\equiv \left\{ \begin{array}{ll}
1+\eta_{-k}-\eta_{2-k} & \\
1+\eta_{1-k}-\eta_{3-k} &
\end{array} \right.
\ (\bmod \ d_2), \ \textrm{if} \ (i, j, l)\ = \left\{ \begin{array}{ll}
(0, 1, 2), (0, 3, 2)& \\
(1, 0, 3), (1, 2, 3)
\end{array} \right.
\end{align*}
For $(i, j, l)\equiv (0, 1, 2)$ or (0, 3, 2), we have

$$-1\equiv \eta_3-\eta_1 \ \textrm{or}\  \eta_1-\eta_3\pmod{d_2}.$$
By Lemma 2.6, $2\equiv 2(\eta_1-\eta_3)^2\equiv G-q,$ $G\equiv q+2$, $q\equiv G^2\equiv q^2+4q+4\pmod {d_2}.$
We get $d_2|\gcd(q^2+3q+4, 2^q+1).$\\
For $(i, j, l)=(1, 0, 3)$ or (1, 2, 3), we have $-1\equiv \pm (\eta_0-\eta_2)\pmod {d_2}$. By Lemma 2.6,
$$2\equiv 2(\eta_0-\eta_1)^2\equiv -G-q, \ G\equiv-(q+2),\ q\equiv G^2\equiv q^2+4q+4 \pmod{d_2}.$$
We also get $d_2|\gcd(q^2+3q+4, 2^q+1).$

In summary, we have proved that $d_1=1$ and $d_2|D=\gcd(q^2+3q+4, 2^q+1).$ This completes the proof of $C_2(\widetilde{S})=\log_2(\frac{2^N-1}{d_1d_2})\geq\log_2(\frac{2^N-1}{D}).$
\end{proof}
\section{ Determination of $C_2(S)$ and $C_2(\widetilde{S})$}
By the definition of 2-adic complexity, $C_2(S)= \log_2(\frac{2^N-1}{d})$ where $d=\gcd(S(2), 2^N-1)$ and $N$ is the period of the sequence $S$. Theorem 3.1 presents nice bounds $\log_2(\frac{2^N-1}{3D})\leq C_2(S)\leq \log_2(\frac{2^N-1}{3})$ for the DHM sequence $S$ where $N=2q$ and $q\equiv 5\pmod 8$ is a prime number, $D=\gcd(q^2+3q+4, 2^q-1).$ In this section, we totally
determine exact value of $C_2(S)$. Firstly, if $D=1$ then $C_2(S)=\log_2(\frac{2^N-1}{3})$. Now we show a necessary and sufficient condition for $D=1$. All arguments belongs to elementary number theory.

 \par\noindent\textbf{Lemma 4.1} Let $q= 8m-3$ be a prime number. Then $D=\gcd(q^2+3q+4, 2^q-1)$ has a prime divisor $l$ if and only if the following two conditions hold.

 (1). $l=2mq+1 (=16m^2-6m+1=\frac{1}{4}(q^2+3q+4)).$

 (2). 2 is a $(2m)$-th power $\pmod l$ which means that there exists $c\in\mathbb{Z}$ such that $2\equiv c^{2m}\pmod l.$
 \begin{proof}
 Remark that if $l=2mq+1,$ then the condition (2) exactly means that $l|(2^q-1).$

 $\Longrightarrow:$ Assume that $l$ is a prime divisor of $D=\gcd(q^2+3q+4, 2^q-1).$ Then $q^2+3q+4\equiv 2^q-1\equiv0\pmod l$. From $2^q\equiv 1\pmod l$ and $q$ is a prime we get $q|l-1$.
 Namely, $l=2qn+1 (n\geq 1).$ Therefore $2nq\equiv -1\pmod l$ and
 $$0\equiv 2n(q^2+3q+4)\equiv -q-3+8n\pmod l.$$
 From $q\geq 5$ and $l=2qn+1\geq 11$ we get
 $$-l=-1-2nq<-q-3+8n<1+2nq=l$$
 which implies that $-q-3+8n=0$. Thus $q=8n-3$ which means that $n=m$ and $l=2mq+1=16m^2-6m+1=\frac{1}{4}(q^2+3q+4).$ Moreover, from $2^q\equiv 1\pmod l$ we know that 2 is a $(\frac{l-1}{q})$-th power$ \pmod l$ where $\frac{l-1}{q}=\frac{16m^2-6m}{8m-3}=2m.$

 $\Longleftarrow:$ Assume that conditions (1) and (2) hold. From condition (2) we know that $2^q \equiv c^{2mq}\equiv c^{l-1}\equiv 1\pmod l.$ Moreover,
 $$q^2+3q+4=(8m-3)^2+3(8m-3)+4=4(16m^2-6m+1)=4l\equiv 0\pmod l.$$
 Therefore $l$ is prime divisor of $\gcd(2^q-1, q^2+3q+4).$
  \end{proof}

  From Theorem 4.1 and Theorem 3.1 we finally determine the 2-adic complexity $C_2(S)$ for the DHM binary sequences as shown in the following result.

  \par\noindent\textbf{Theorem 4.2} Let $q\equiv 5\pmod 8$ be a prime number, $q=s^2+4t^2, s\equiv 1\pmod 4$, $S=\{s_\lambda\}_{\lambda=0}^\infty$ be the DHM binary sequence with period $N=2q$ defined
  by Definition 2.1 where one of the following conditions holds

  (I,1) $t=1$ and $(i, j, l)=(0, 1, 3), (0, 2, 1);$ or

  (I,2) $s=1$ and $(i, j, l)=(1, 0, 3), (0, 1, 2).$

  Let $D=\frac{q^2+3q+4}{4}$ and $C_2(S)$ be the 2-adic complexity of the DHM binary sequence $S$. Then

  $$\ C_2(S)=\log_2(\frac{2^N-1}{3}) \ \textrm{or} \ \log_2(\frac{2^N-1}{3D}). \ \  \textrm{Moreover, }\ C_2(S)=\log_2(\frac{2^N-1}{3D})$$
  if and only if $D$ is a prime number, $2^q\equiv 1\bmod D$ and $D|S(2)=\sum_{\lambda=0}^{N-1}s_\lambda 2^\lambda$.

  Remark (1). By Lemma 4.1, we need to check if 2 is a $(2m)$-th power $\pmod D$. A series of criteria for 2 being an $n$-th power $\pmod D$ for several $n$ are given in Book [1], which are helpful to simplify computation.

  (2). From Theorem 4.2 we know that if $q\equiv 5\pmod 8$ is a prime and $M_q=2^q-1$ is also a prime (called Mersenne prime), then $C_2(S)=\log_2(\frac{4^q-1}{3})$. The table 2.9 in the Book [7], Section 2.7 presented all Mersenne primes $M_q$ for $q\leq 7\times 10^6$.  There are 12  Mersenne primes $M_q$ with $q\equiv 5\pmod 8: q$=5, 13, 61, 4253, 9941, 11213, 21701, 1398269, 2976221, 13466917, 77232917, and 82589933.

  Similarly we have the following result.

 \par\noindent\textbf{Lemma 4.3} Let $q=8m-3$ be a prime number, $D=\gcd(q^2+3q+4, 2^q+1).$ Then $D>1$ if and only if

 (1). $D=2mq+1 (=16m^2-6m+1=\frac{1}{4}(q^2+3q+4))$ and $D$ is a prime.

 (2). $2^q\equiv -1\pmod D$ which means that 2 is an $m$-th power $\pmod D$.
 \begin{proof}($\Rightarrow:$) Let $D> 1$ and $l$ be a prime divisor of $D$. Then $2^q\equiv -1$ and $2^{2q}\equiv 1\pmod l$. Since $3\nmid q^2+3q+4$
 we know that $l\neq 3$, and then the order of $2\pmod l$ is $2q$. Therefore $2q|l-1$ and $l=2nq+1$ $(n\geq 1)$. Then $2nq\equiv -1\bmod l$ and
 $$0\equiv 2n(q^2+3q+4)\equiv -q-3+8n \pmod l.$$
 From $q\geq 5$ and $l=2nq+1\geq 11$ we get
 $$-l=-1-2nq<-q-3+8n<1+2nq=l.$$
 Therefore $-q-3+8n=0,$ $q=8n-3$ so that $n=m.$ Namely,
 $$l=2mq+1=D=\frac{1}{4}(q^2+3q+4)$$
 which means that $D$ is a prime. The condition (2) comes from $2^q\equiv -1\pmod l$ and $l=D$.

 $(\Leftarrow:)$ If the conditions (1) and (2) holds, then $\gcd(q^2+3q+4, 2^q+1)=\gcd(4D, 2^q+1)=D.$
 \end{proof}

 From Theorem 3.2 and Lemma 4.3 we finally determine the 2-adic complexity $C_2(\widetilde{S})$ for the DHM sequences $\widetilde{S}$ as shown in the following result.

\textbf{Theorem 4.4} Let $q\equiv 5\pmod 8$ be a prime number, $q=s^2+4t^2$, $s\equiv 1\pmod 4$, $\widetilde{S}=\widetilde{S}(i, j, l)$ be the DHM binary sequence with period $N=2q$ defined by Definition 2.1, where one of the following conditions holds

(II, 1) $t=1, (i, j, l)=(0, 1, 3), (0, 2, 3), (1, 2, 0), (1, 3, 0)$ or

(II, 2) $s=1, (i, j, l)=(0, 1, 2), (0, 3, 2), (1, 0, 3), (1, 2, 3)$.

Let $D=\frac{1}{4}(q^2+3q+4)$. Then $C_2(\widetilde{S})=\log_2(2^N-1)$ or $\log_2(\frac{2^N-1}{D})$. Moreover, $C_2(\widetilde{S})=\log_2(\frac{2^N-1}{D})$ if and only if $D$ is a prime number, $2^q\equiv -1\pmod D$ and $D|\widetilde{S}(2).$

 At the end of this section we present a small example to show that $C_2(\widetilde{S})$ can be less than $\log_2(2^N-1).$

   \par\noindent\textbf{Example} Let $q=5=1+4, s=1, \mathbb{F}_5^{\ast}=\langle3\rangle$. The cyclotomic classes of order 4 in $\mathbb{F}_5$
   are
   $$D_0=\{1\}, D_1=\{3\}, D_2=\{3^2=4\}, D_3=\{3^3=2\}.$$
   Thus $2\in D_k, k=3,$ and
   $$\eta_0\equiv4, \eta_1\equiv4^3, \eta_2\equiv 4^4, \eta_3\equiv 4^2\pmod {2^N-1}, N=2q=10.$$
   On the other hand, $D=\frac{1}{4}(q^2+3q+4)=11$ is a prime number and $2^q\equiv 2^5\equiv -1\pmod {11}$. \\
   For $(i, j, l)=(1, 0, 3)$ and (1, 2, 3),
   $$\widetilde{S}(2)\equiv 1+\eta_{i-k}-\eta_{l-k}\equiv 1+\eta_2-\eta_0\equiv 1+4^4-4\equiv 1+3-4\equiv 0\pmod {11}.$$
   By Theorem 4.4, for $\widetilde{S}=\widetilde{S}(1, 0, 3)$ and $\widetilde{S}(1, 2, 3)$ we have $C_2(\widetilde{S})=\log_2(\frac{2^{10}-1}{11})=\log_293.$\\
   For $(i, j, l)=(0, 1, 2)$ and (0, 3, 2), $\widetilde{S}(2)\equiv 1+\eta_1-\eta_3\equiv 1+9-5\not\equiv 0\pmod {11}.$ We get
   $C_2(\widetilde{S})=\log_2(2^{10}-1).$ In fact, the DHM sequence $\widetilde{S}=\widetilde{S}(i, j, l)$ is

 $$\widetilde{S}=(1 \ 1 \ 0 \ 0 \ 0 \ 0 \ 1 \ 1 \ 1 \ 0\ldots), \textrm{for} \ (i, j, l)=(1, 0, 3)$$

 $$\widetilde{S}(2)=1+2+2^6+2^7+2^8\equiv 1+2+9+7+3\equiv 0\pmod {11}$$

 and

$$\widetilde{S}=(1 \ 0 \ 0 \ 0 \ 1 \ 0 \ 0 \ 1 \ 1 \ 1\ldots), \textrm{for} \ (i, j, l)=(1, 2, 3)$$

 $$\widetilde{S}(2)=1+2^4+2^7+2^8+2^9\equiv 1+5+7+3+6\equiv 0\pmod {11}$$

\newcolumntype{d}{D{.}{.}{2}}
\begin{tabular}{|c|*{4}{d|}}
\hline
 $\lambda$ & 0 \ \ \ 1 \  \ \ 2  \ \ \ 3 \ \ \ 4 \ \ \ 5\ \ \ 6 \ \ \ 7  \ \ \ 8  \ \ \ 9 \\ \hline
$\lambda\pmod 2$ & 0 \ \ \ 1\  \ \  0 \ \ \ 1 \ \ \ 0 \ \ \ 1\ \ \ 0 \ \ \ 1  \ \ \ 0  \ \ \ 1 \\ \hline
$\lambda\pmod 5$ & 0 \ \ \ 1\  \ \ 2 \ \ \ 3 \ \ \ 4 \ \ \ 0\ \ \ 1 \ \ \ 2  \ \ \ 3  \ \ \ 4 \\ \hline
$\widetilde{s}_\lambda((i, j, l)=(1, 0, 3))$ &  1 \ \ \ 1 \  \ \ 0 \ \ \ 0 \ \ \ 0 \ \ \ 0\ \ \ 1 \ \ \ 1  \ \ \ 1 \ \ \ 0\\ \hline
$\widetilde{s}_\lambda((i, j, l)=(1, 2, 3))$ &  1 \ \ \ 0 \  \ \ 0 \ \ \ 0 \ \ \ 1 \ \ \ 0\ \ \ 0 \ \ \ 1  \ \ \ 1 \ \ \ 1\\ \hline
\end{tabular}

\end{document}